\documentclass[12pt,a4paper]{article}
\usepackage{amsfonts}
\usepackage{amssymb}
\usepackage{bm}

\advance\voffset by -2.0cm \advance\hoffset by -1.25cm
\def\AEF{A.E. Faraggi}

\def\PLB#1#2#3{{\it Phys.\ Lett.}\/ {\bf B#1} (#2) #3}
\def\PLA#1#2#3{{\it Phys.\ Lett.}\/ {\bf A#1} (#2) #3}
\def\PRD#1#2#3{{\it Phys.\ Rev.}\/ {\bf D#1} (#2) #3}

\def\IJMP#1#2#3{{\it Int.\ J.\ Mod.\ Phys.}\/ {\bf A#1} (#2) #3}

\def\EJP#1#2#3{{\it Eur.\ Phys.\ Jour.}\/ {\bf C#1} (#2) #3}

\def\JCA#1#2#3{{\it JCAP}\/ {\bf #1} (#2) #3}
\def\etal{{\it et al\/}}

\usepackage{amsmath}
\usepackage{amssymb}
\usepackage{amsthm}
\usepackage{amsfonts}

\numberwithin{equation}{section}

\theoremstyle{definition}

\newcommand{\RR}{\mathbb{R}} 

\newcommand{\beq}{\begin{equation}}
\newcommand{\eeq}{\end{equation}}
\newcommand{\beqa}{\begin{eqnarray}}
\newcommand{\beqn}{\begin{eqnarray}}
\newcommand{\eeqn}{\end{eqnarray}}
\newcommand{\eeqa}{\end{eqnarray}}

\def\1{\frak 1}
\def\2{\frak 2}
\def\3{\frak 3}

\newlength{\oldcolsep}

\begin{document}


\title{The Quantum Closet}
\author{Alon E. Faraggi
}\date{}

\maketitle

\begin{center}{Department of Mathematical Sciences\\
               University of Liverpool, Liverpool L69 7ZL, UK }
\end{center}

\bigskip

\bigskip

\begin{abstract}

The equivalence postulate approach to quantum mechanics entails a derivation 
of quantum mechanics from a fundamental geometrical principle. Underlying the
formalism there exists a basic cocycle condition, which is invariant under
$D$--dimensional
finite M\"obius transformations. The invariance of the cocycle
condition under finite 
M\"obius transformations implies that space is compact. Additionally,
it implies 
energy quantisation and the undefinability of
quantum trajectories. 
I argue that the decompactification limit coincides with the 
classical limit. 
Evidence for the compactness of the universe may 
exist in the Cosmic Microwave Background Radiation.

\end{abstract}

\noindent

\newpage
\setcounter{footnote}{0}
\renewcommand{\thefootnote}{\arabic{footnote}}

\section{Introduction}

The synthesis of quantum mechanics and general relativity continues to pose 
an important challenge in the basic understanding of physics. While quantum 
mechanics accounts with astonishing success for physical observations at 
the smallest distance scales, general relativity accomplishes a similar
feat at the largest. Yet these two mathematical modellings of the observed
data are mutually incompatible. This is seen most clearly in relation to the 
vacuum. The first predicts a value that is off by orders of magnitude from
the observed value, which is determined by using the second. To date there is
no solution to this problem. 
In view of this calamity it seems prudent to  explore the
foundations of each of 
these theories, and the 
fundamental principles that underly them.
General relativity follows from a basic geometrical principle, the equivalence 
principle, whereas the basic tenant of quantum mechanics is the 
probability interpretation of the wave function.

The question arises whether quantum mechanics can follow from a basic 
geometrical principle, akin to the geometrical principle that
underlies relativity. 
Starting in \cite{fm} we embarked on a rigorous derivation of 
quantum mechanics from a geometrical principle. The equivalence postulate
of quantum mechanics hypothesises that any two physical states can be
connected by a coordinate transformation. This includes states which arise
under different potentials. In particular, any state may be transformed 
so as to correspond to that of a free particle at rest.
This bears close resemblance
to Einstein's equivalence principle that underlies general relativity with an 
important caveat. While in the case of Einstein's equivalence principle 
it is the gravitational field which is ``locally balanced'' by a coordinate
transformation, in the equivalence postulate approach to quantum mechanics
it is an arbitrary external potential which is ``globally balanced'' by 
a coordinate transformation.
%
The equivalence 
postulate of quantum mechanics is naturally formulated in the framework
of Hamilton--Jacobi theory. 

The implementation of the equivalence postulate in the context of the 
Hamilton--Jacobi theory yields a Quantum Hamilton--Jacobi equation.
The Classical 
Hamilton--Jacobi Equation is obtained by requiring the existence of 
a canonical transformation from one set of phase space variables to a second
set of phase space variable such that the Hamiltonian is mapped to a trivial 
Hamiltonian. Consequently, the new phase--space variable are 
constants of the motion, {\it i.e.}
\beq
H(q,p)~~\longrightarrow~~ { K}({ Q},{P})\equiv0~~~~~~~~
\Longrightarrow~~~~~\dot{  Q}={\partial{K}\over
{\partial{P}}}\equiv 0~,~\dot{P}=
-{\partial {K}\over{\partial Q}}\equiv 0. ~~~~~~~~~
\label{canonicaltrans}
\eeq
The solution to this problem is given by the Classical
Hamilton--Jacobi equation
(CHJE). 
Since the transformations are canonical the phase space variables are taken as 
independent variables and their functional dependence is only extracted 
from the solution of the CHJE via the functional relation 
\beq
p={{\partial S(q)}\over {\partial q}}, 
\label{peqdsdq}
\eeq
where $S(q)$ is Hamilton's principal
function. The fundamental uncertainty relations
of quantum mechanics imply that the phase--space variables are not independent. 
The equivalence postulate of quantum mechanics therefore requires the existence 
of trivialising coordinate transformations for any physical system,  but the 
phase--space variables are not independent in the application of the
trivialising transformations. They are related by a generating function,
via (\ref{peqdsdq}),
which transforms as a scalar function under the transformations. 
That is, 
\beq
(~q~,~S_0(q)~,~p~=~{{\partial S_0}\over{\partial q}})~~
\longrightarrow~~
(~{q}^v~,~{S_0^v}( q^v)~,~{p^v}~=~
{{\partial{S}_0^v}\over{\partial{ q^v}}}~), 
\label{vmap}
\eeq
where $S_0(q)$ is the generating function in the stationary case.
It is instrumental
to study the stationary case in order to see the symmetry 
structure that underlies quantum mechanics. 
The consistency of the equivalence hypothesis implies that the Hamilton--Jacobi equation retains its
form under coordinate transformations. However, this cannot be implemented in
classical mechanics.
The CSHJE for a particle  moving under the influence of a velocity independent
potential $V(q)$ is given by
\begin{equation}
\label{cshje} \frac{1}{2m}\sum_{i=1}^N\left(\frac{\partial
S}{\partial q_i}\right)^2 + \mathcal{W}(q) = 0,
\end{equation}
where $\mathcal{W}(q)\equiv V(q)-E$. Under a change of
coordinates $q\rightarrow q^v$ we have (by (\ref{vmap}))
\begin{equation}
\label{jac} \frac{\partial S^v(q^v)}{\partial q^v_j} =
\frac{\partial S(q)}{\partial q^v_j} = \sum_i \frac{\partial 
S(q)}{\partial q_i}\frac{\partial q_i}{\partial q^v_j},
\end{equation}
which we can write as
$\bm{p}^v=\mathbf{J}^v\bm{p}, $
where
$J^v_{ij}=\frac{\partial q_i}{\partial q^v_j}$ is the Jacobian
matrix connecting the coordinate systems $q$ and $q^v$, and where,
$p_i=\frac{\partial S}{\partial q_i}$. Then
\begin{equation}
\nonumber \sum_j \left(\frac{\partial S^v}{\partial
q^v_j}\right)^2=|\bm{p}^v|^2\\
=\left(\frac{|\bm{p}^v|^2}{|\bm{p}|^2}\right)|\bm{p}|^2\\
= (p^v|p)|\bm{p}|^2,\label{p2def} 
\end{equation}
where we have defined
\begin{equation}
 (p^v|p) \equiv
\frac{|\bm{p}^v|^2}{|\bm{p}|^2}
= \frac{{\bm{p}^v}^\mathsf{T}\bm{p}^v}{\bm{p}^\mathsf{T}\bm{p}}
= \frac{\bm{p}^\mathsf{T}
{\mathbf{J}^v}^\mathsf{T} \mathbf{J}^v \bm{p}}{\bm{p}^\mathsf{T}
\bm{p}}.\label{trafact}
\end{equation}
It is seen that the first term in eq. (\ref{cshje}) transforms
as a quadratic differential 
under the $v$--map eq. (\ref{vmap}). Since $S_0^v(q^v)$ must
satisfy the CSHJE,  
covariance of the HJ equation under the $v$--transformations
implies that the second term in eq. (\ref{cshje}) transforms
as a quadratic differential.
That is 
\begin{equation}
 \mathcal{W}^v(q^v) = (p^v|p)\mathcal{W}(q).
\label{Wtran}
\end{equation}
In particular, for the $\mathcal{W}^0(q^0)\equiv 0$ state we have,  
\begin{equation}
 \mathcal{W}^0(q^0)~\longrightarrow ~\mathcal{W}^v(q^v)~=~
(p^v|p^0)\mathcal{W}^0(q^0)~=~0.
\label{W0}
\end{equation}
This means that $\mathcal{W}^0$ is a \emph{fixed point} under $v$-maps,
{\it i..e.} it
cannot be connected to other states. Hence, 
we conclude that the equivalence postulate cannot be implemented consistently 
in classical mechanics.

\section{The cocycle condition}\label{coco}

Consistent implementation of the equivalence postulate necessitates
the modification of classical mechanics, which entails adding a yet to be determined function, $\mathcal{Q}(q)$, 
to the CSHJE.
This augmentation produces the Quantum Stationary 
Hamilton--Jacobi Equation (QSHJE)
\beq
{1\over{2m}}\left({{\partial S(q)}\over {\partial q}}\right)^2+ 
\mathcal{W}(q)+\mathcal{Q}(q)=0~,
\label{qshje}
\eeq
where $\mathcal{W}(q)=V(q)-E$.
It is noted that the combination
$\mathcal{W}(q)+\mathcal{Q}(q)$ transforms 
as a quadratic differential under coordinate transformations, 
whereas each of the functions $\mathcal{W}(q)$ and $\mathcal{Q}(q)$ 
transforms as a quadratic differential up to an additive term,
{\it i.e.} under $q^a\rightarrow {q}^v(q)$ we have, 
\beqn
\mathcal{W}^a(q^a)\rightarrow  \mathcal{W}^v({q^v}) & = &
              \left( p^v\vert p^a\right)
                        \mathcal{W}^a({q^a}) + (q^a;{q}^v)\nonumber\\
\mathcal{Q}^a(q^a)\rightarrow  \mathcal{Q}^v({q}^v) & = &
              \left( p^v\vert p^a\right)
                        \mathcal{Q}^a({q}^a) - (q^a;{q^v}).\nonumber
\eeqn
and 
\beq
\left(
\mathcal{W}(q^a)+ \mathcal{Q}(q^a)
\right)\rightarrow  
\left(
\mathcal{W}^v({q^v})+ \mathcal{Q}^v(q^v)
\right)= 
\left( p^v\vert p^a \right)                        
\left(
\mathcal{W}^a(q^a)+\mathcal{Q}^a(q^a)
\right)\nonumber
\eeq 
All physical states with a non--trivial 
$\mathcal{W}(q)$ then arise 
from the inhomogeneous part in the transformation of the 
trivial state $\mathcal{W}^0(q^0)\equiv0$, 
{\it i.e.} $\mathcal{W}(q)= (q^0; q)$. 
Considering the transformation $q^a\rightarrow q^b\rightarrow q^c$
versus $q^a\rightarrow q^c$ gives rise to the cocycle
condition on the inhomogeneous term 
\beq
(q^a;q^c)= \left( p^c\vert p^b\right)
           \left[ (q^a;q^b) - (q^c;q^b)\right]. 
\label{cocycle}
\eeq
The {\it cocycle condition} eq. (\ref{cocycle}) embodies the essence of 
quantum mechanics in 
the equivalence postulate approach. 
Furthermore, it reveals the basic symmetry properties that 
underly quantum mechanics. It is proven \cite{fm,bfm}
that the cocycle condition is invariant under
$D$--dimensional M\"obius transformations, which include translations,
dilatations, rotations 
and, most crucially, inversions, or reflections, in the unit sphere.
The M\"obius transformations are, hence,
defined on the compactified space $\hat\RR^D=\RR^D\cup\{\infty\}$.
Whereas translations, dilatation 
and rotations map $\infty$ to itself, inversions exchange
$0 \leftrightarrow\infty$. 
We argue that energy quantisation and the existence of a fundamental length
scale in the formalism, together with the invariance of the
cocycle condition eq. (\ref{cocycle})
under the M\"obius group $M(\hat\RR^D)$ of transformations,
implies that space is compact.
The more general situation may be considered in the
decompactification limit. 

The cocycle condition fixes the functional form of the quantum potential
$\mathcal{Q}(q)$. 
In one dimension the cocycle condition (\ref{cocycle})
fixes the inhomogeneous 
term $$(q^a; q^b)=-\beta \{ q^a, q^b\}/{4m},$$ where
$\{f,q\} =f^{\prime\prime\prime}/f^\prime-3(f^{\prime\prime}/f^\prime)^2/2$ 
the Schwarzian derivative and $\beta$ is a constant
with the dimension of an action. In one dimension the
Quantum Hamilton--Jacobi equation is given in terms of a basic Schwarzian
identity, 
\begin{equation}
\left({{\partial S(q)}\over {\partial q}}\right)^2=   {\beta^2\over 2}
\left(\left\{{\rm e}^{{{2i}\over\beta}{S}},q\right\}-\left\{S,q\right\}\right)
\label{schwarzianid}
\end{equation}
Making the identification
\beq
\mathcal{W}(q)= V(q)-E =
-{\beta^2\over {4m}}\left\{{\rm e}^{{{(i2S_0)}\over \beta}},q\right\},
\label{wqeqvqminuse}
\eeq
and
\beq
\mathcal{Q}(q) = {\beta^2\over {4m}}\left\{S_0,q\right\},
\label{qq}
\eeq
we have that $S_0$ is the solution of the Quantum Stationary
Hamilton--Jacobi
equation (QSHJE),
\beq
{1\over {2m}}\left({{\partial S_0}\over {\partial q}}\right)^2 +
V(q) - E + {\hbar^2\over{4m}}
\left\{S_0,q\right\} = 0 . 
\label{qshje2}
\eeq
The Schwarzian identity, eq. (\ref{schwarzianid}), 
is generalised in higher dimensions by the basic quadratic identity
\beq 
\alpha^2(\nabla S_0)^2=
{\Delta(R{\rm e}^{\alpha S_0})\over R{\rm e}^{\alpha S_0}}-
{\Delta R\over R}-{\alpha\over R^2}\nabla\cdot(R^2\nabla S_0), 
\label{hdid}
\eeq
which holds for any constant $\alpha$
and any functions $R$ and $S_0$. Then, if $R$ 
satisfies the continuity equation 
$\nabla\cdot(R^2\nabla S_0)=0,$
and setting $\alpha=i/\hbar$, we have 
\beq
{1\over2m}(\nabla S_0)^2=
-{\hbar^2\over2m}{\Delta(R{\rm e}^{{i\over 
\hbar} S_0})\over R{\rm e}^{{i\over\hbar}S_0}}+
{\hbar^2\over2m}{\Delta R\over R}. 
\label{id2}
\eeq 
In analogy with the one dimensional
case we make identifications, 
\beqn
\mathcal{W}(q)=V(q)-E& = &
{\hbar^2\over2m}{\Delta(R{\rm e}^{{i\over\hbar}S_0})\over 
R{\rm e}^{{i\over \hbar}S_0}}, 
\label{hdwq}\\
\mathcal{Q}(q)& =& 
-{\hbar^2\over2m}{\Delta R\over R}. 
\label{id3}
\eeqn
Eq. (\ref{hdwq}) implies the $D$--dimensional Schr\"odinger equation
\beq
\left[-{\hbar^2\over2m}{\Delta}+V(q)\right]\Psi=E\Psi. 
\label{hdschroedingerequation}
\eeq
and the general solution
\beq
\Psi= R(q) 
\left( A {\rm e}^{{i\over \hbar} S_0} 
+                                                   
       B {\rm e}^{-{i\over \hbar} S_0}\right).
\label{hdwavefunction}
\eeq 
We note that consistency of the equivalence postulate formalism
necessitates that the
two solutions of the 
second order Schr\"odinger equation are retained. This is reminiscent
of relativistic 
quantum mechanics in which both the positive and negative energy
solutions are retained. 
We can replace the gradient in eq. (\ref{hdid}) by a four vector
derivative in Minkowski space. 
This produces the generalisation of the formalism to the relativistic
case and the Schr\"odinger 
equation, eq. (\ref{hdschroedingerequation}), is replaced by the
Klein--Gordon equation. 
The time--dependent Schr\"odinger equation arises in the limit
$c\rightarrow\infty$. 
Similarly, the cocycle condition eq. (\ref{cocycle}) generalises to
Minkowski space
by replacing the Euclidean metric with the Minkowski metric. 
It is important to emphasize that the equivalence postulate
approach to quantum
mechanics does not represent a modification or interpretation
of quantum mechanics
but its derivation from a basic geometrical principle. As such it
reveals the geometrical
structures underlying quantum mechanics and in that respect provides
an intrinsic framework
to explore the quantum space--time. It is further noted that the
cocycle condition, eq. (\ref{cocycle}),
is completely universal. Hence, its generalisation to curved
space provides a background
independent approach to quantum gravity. In this respect
the equivalence postulate approach
reveals the interplay between quantum variables, encoded
$R(q)$ and $S(q)$, versus the classical
background parameters. For example, in ref. \cite{eqtp}
we showed that the QHJE
does not admit a consistent time parameterisation of
quantum trajectories. In this respect,
therefore, time cannot be defined as a quantum observable,
but is merely a classical background
parameter. Generalising this observation to relativistic
space-time entails that space--time
cannot be consistently defined as a quantum observable.
Instead, the quantum data 
is encoded in the cocycle condition and the corresponding
quadric identity in the
relevant domain, {\it i.e.} in curved space--time.
In this respect, we note that the inhomogeneous
term can be written in the general form \cite{bfm}, 

\beq
(q^a;q^b)=(p^b|p^a)
\mathcal{Q}^a(q^a)-\mathcal{Q}^b(q^b)=-{\hbar^2\over2m}\left[(p^b|p^a) 
{\Delta^aR^a\over R^a}-{\Delta^bR^b\over R^b}\right],  
\label{qaqbrarb}
\eeq
which shows how the information on the inhomogeneous term 
is encoded in the functions 
$R(q)$ and $S(q)$.

\section{The quantum closet}\label{tp}

The invariance of the cocycle condition under M\"obius transformations
implies that space is compact. 
Let us gather the evidence for this claim. In the one dimensional
case we see from eq. (\ref{hdwq})
that the QSHJE is equivalent to the equation
$\{w,q\}=-4m(V(q)-E)/\hbar^2$ where $w$ is the ratio of the two
solutions of the 
Schr\"odinger equation. It follows from the M\"obius invariance of
the cocycle condition that
$w\ne const$, $w\in C^2({\hat \RR})$ with
$w^{\prime\prime}$ differentiable on $\RR$, 
where ${\hat \RR}=\RR\cup\{\infty\}$, 
and
\beq
w(-\infty)=\left\{
\begin{array}{ll}
+w(+\infty) \, & {\rm if} \,\,\, w(-\infty)\ne \pm \infty \ ,\\
-w(+\infty) \, & {\rm if} \,\,\, w(-\infty)=\pm\infty \ .
\end{array}\right.
\label{winfty}
\eeq
Furthermore, denoting by $q_-$ ($q_+$) the lowest (highest)
$q$ for which $V(q)-E$ changes sign, we prove the general theorem \cite{fm},

\noindent
{\it If}
\beq
V(q)-E\geq\left\{
\begin{array}
{ll}P_-^2 >0 \ , & q < q_- \ ,\\ 
           P_+^2 >0 \ , & q > q_+ \ ,
\end{array}\right.
\label{boundingpotential}
\eeq
{\it then $w=\psi^D/\psi$ is a
local self--homeomorphism of
$\hat\RR$ iff the
Schr\"odinger equation has an
$L^2(\RR)$ solution.}

\noindent
Since the QSHJE is defined if and only if $w$ is a local self--homeomorphism
of $\hat\RR$, this theorem implies that energy quantisation {\it directly}
follows from the geometrical gluing conditions of $w$ at $q=\pm\infty$,
as implied by the equivalence postulate, which in turn imply that the one
dimensional space is compact. In turn the compactness of space implies that
the energy of the free quantum particle is quantised and that time
parameterisation of trajectories  is ill defined either via
Bohm--de Broglie mechanical definition, or via Floyd's definition
by using Jacobi's theorem \cite{floyd}.  The M\"obius invariance of the
cocycle condition in $D$ dimensions then implies that 
the $D$ dimensional space is compact. 

Generalisation of the cocycle condition to curved space
suggests a background independent approach to quantum
gravity. The connection with gravity and 
with an internal structure of elementary particles is
implied due to the existence of an intrinsic fundamental length 
scale in the formalism, and the association of the quantum
potential, $\mathcal{Q}(q)$, with a curvature term 
\cite{fm,marcosgravity,epde}.
%
%
To see the origin of that we can again examine the stationary one
dimensional case
with $\mathcal{W}^0(q^0)\equiv0$. In this case 
the Schr\"odinger equation takes the form
$${\partial^2\over{\partial q}^2}\psi=0,$$
with the two linearly independent solutions being $\psi^D=q^0$ and 
$\psi=const$. Consistency of the equivalence postulate 
dictates that both solutions of the Schr\"odinger equation
must be retained. The solution of the corresponding 
QHJE is given by \cite{fm}  
$$
{\rm e}^{{2i\over\hbar}S_0^0}={\rm e}^{i\alpha}
{{q^0+i{\bar\ell}_0}\over{q^0-i\ell_0}}, 
$$
where $\ell_0$ is a constant with the dimension of 
length \cite{fm}, 
and the conjugate momentum 
$p_0=\partial_{q^0} S_0^0$ takes the form
\beq
p_0=\pm{{\hbar (\ell_0+{\bar\ell}_0) }\over {2\vert q^0- i\ell_0\vert^2}}.
\label{p00}
\eeq
It is seen that $p_0$ vanishes only for $q^0\rightarrow\pm \infty$.
The requirement that in the classical limit
$\lim_{\hbar\rightarrow0}p_0=0$ implies that we can set \cite{fm}
\beq
{\rm Re}\,\ell_0=\lambda_p= \sqrt{{\hbar G}\over c^3}, 
\label{settingell0}
\eeq
{\it i.e.} we identify ${\rm Re}\,\ell_0$ with the Planck length.
The interpretation of the quantum potential as a curvature term 
\cite{fm,epde} implies that elementary particles possess 
an internal structure, {\it i.e.} points do not have curvatures. 
This suggests possible connection with theories of extended objects.

If the universe is compact it would imply the existence of an
intrinsic energy scale reminiscent of the Casimir effect. Taking the
present size of the observable universe would imply a very small 
energy scale, which is essentially unobservable \cite{epde}. 
However, given the indication of a larger energy scale in the Cosmic
Microwave Background (CMB) Radiation suggests the possibility of 
observing the imprints of compactness of the universe in the CMB 
in the current \cite{planck} or future CMB observatories. 
Indeed, the possibility of signatures of a non--trivial topology
in the CMB has been of recent interest \cite{topcmb}. Additional 
experimental evidence for the equivalence postulate approach
to quantum mechanics may arise from modifications of the
relativistic energy--momentum relation \cite{mdr}, which affects the 
propagation of light from gamma ray bursts \cite{aemns}. 

\section{The decompactification limit}

The M\"obius invariance of cocycle condition may only be implemented 
if space is compact. We may contemplate that the decompactification limit
represents the case when the spectrum of the free quantum particle becomes
continuous. In that case time parameterisation of quantum trajectories
is consistent with the definition of time by using Jacobi's theorem
\cite{floyd,fm,eqtp}. However, I argue that the decompactification 
limit in fact concides with the classical limit. To see that this 
may be the case we examine again the case of the free particle 
in one dimension. 
The quantum potential associated with the state
$W^0\equiv0$ is given by
\beq
\mathcal{Q}^0={\hbar^2\over4m}\{S_0^0,q^0\}= 
-{{\hbar^2 ({\rm Re}\,\ell_0)^2}\over2m}
{1\over{\vert q^0-i\ell_0\vert^4}}.
\label{Q0pot}
\eeq
We note that the limit $q^0\rightarrow\infty$ concides with the 
limit $\mathcal{Q}^0\rightarrow 0$, {\it i.e.} with the classical limit.

\section{Conclusions}\label{conclude}

Heisenberg's uncertainty principle mandates that the phase--space variables
cannot be treated as independent variables. The classical Hamilton--Jacobi
trivialising transformations are
in direct conflict with this fact. Reconciling the Hamilton--Jacobi 
theory to quantum mechanics leads to the quantum Hamilton--Jacobi 
equation (QHJE). In turn, the QHJE implies a basic cocycle condition
that underlies quantum mechanics. The cocycle condition holds in 
any background and provides a framework for the background
independent formulation of quantum gravity. The cocycle condition
is invariant under $D$--dimensional finite M\"obius transformations
with respect to the Euclidean or Minkowski metrics. Its invariance under
$D$--dimensional M\"obius transformations implies that space
is compact, which may have an imprint in the cosmic microwave 
background radiation. 

\section*{Acknowledgments}

I thank Marco Matone for discussions and Subir Sarkar and 
Theoretical Physics Department at the University of Oxford for hospitality.
This work is supported in part by the STFC (PP/D000416/1).

\bibliography{apssamp}

\end{document}